\documentclass[namedreferences]{SolarPhysics}
\usepackage[optionalrh]{spr-sola-addons} % For Solar Physics
\usepackage{graphicx}        % For eps figures, newer & more powerfull
\usepackage{color}           % For color text: \color command
\usepackage{url}             % For breaking URLs easily trough lines
\newcommand{\etal}{{\it et al.}}

\newcommand{\aap}{    {\it Astron. Astrophys.}}

\newcommand{\apj}{    {\it Astrophys. J.}}

\newcommand{\apss}{   {\it Astrophys. Spa. Sci.}}

\newcommand{\jgr}{    {\it J. Geophys. Res.}}

\newcommand{\solphys}{{\it Solar Phys.}}

\begin{document}
\begin{article}
\begin{opening}
\title{Long-term Variations of Solar Differential Rotation and Sunspot Activity: Revisited}
\author{K.J. ~\surname{Li}$^{1,2}$\sep
        W. ~\surname{Feng}$^{3}$\sep
        X.J. ~\surname{Shi}$^{1,4}$\sep
        J.L. ~\surname{Xie}$^{1,4}$ \sep
        P.X. ~\surname{Gao}$^{1}$ \sep
        H.F. ~\surname{Liang}$^{5}$
       }
\runningauthor{ K.J. Li \etal}

\runningtitle{Differential Rotation and Spot Activity}
\institute{$^{1}$National Astronomical Observatories/Yunnan
Observatory, CAS, Kunming 650011, China;
email: \url{lkj@mail.ynao.ac.cn} \\
$^{2}$Key Laboratory of Solar Activity, National Astronomical
Observatories, CAS, Beijing 100012, China \\
$^{3}$ Research Center of Analysis and Measurement,
Kunming University of Science and Technology, Kunming 650093, China \\
$^{4}$ Graduate School, CAS, Beijing 100863, China\\
$^{5}$Department of Physics, Yunnan Normal University, Kunming 650093,
 China }
%\date{Received: 14 January 2008/Accepted: 21 October 2008}

\begin{abstract}
Long-term variations of solar differential rotation and sunspot activity are investigated through re-analyzing the data on parameters of the differential rotation law obtained by  Makarov,  Tlatov, and  Callebaut (1997),
Javaraiah, Bertello, and  Ulrich (2005a, b), and Javaraiah et al. (2009).
Our results indicate that the solar surface rotation rate at the Equator
(indicated by the A parameter of the standard solar rotation law) shows a
secular decrease since cycle 12 onwards, given by about $1\,-\,1.5\times10^{-3}$($deg\ day^{-1} year^{-1}$). The B parameter of the standard differential rotation
law seems to also show a secular decrease since cycle 12 onwards, but of
weak statistical significance. The rotation rate averaged on latitudes
($0^{o}\,--\,40^{o}$) does not show a secular trend of statistical significance.
Moreover, the average sunspot area shows a secular increase of statistical
significance since cycle 12 onwards, while a negative correlation is found
between the level of sunspot activity (indicated by the average sunspot area)
 and the solar equatorial rotation in the long run.

\end{abstract}
\end{opening}
%-------------------------------------------------
\section{Introduction}\label{S-Introduction}
Solar differential rotation was discovered in 1630
when Christoph Scheiner first noticed that the solar equatorial region should rotate
faster than higher latitude regions: 26 days at the solar Equator and 30 days at $60^{\circ}$ latitude
of the Sun (Le Mouel, Shnirman, and Blanter, 2007).  Since the basic work of Carrington on solar rotation, many studies
have been made to investigate the solar differential rotation using different data and different methods, which can be divided into  two kinds: the tracer method and the spectroscopic
method (Balthasar and Wohl, 1980; Ternullo,  Zappala, and Zuccarello, 1981;
Balthasar, Vazquez, and Wohl, 1980; Arevalo et al., 1982;
Gilman and Howard, 1984; Howard, Gilman, and Gilman, 1984; Balthasar et al., 1986;
Sheeley, Nash, and Wang, 1987; Tuominen, and Virtanen, 1987;
 Nash, Sheeley, and Wang, 1988; Zappala, and Zuccarello, 1989; Snodgrass and Ulrich, 1990;
  Brajsa et al. ,1992; Japaridze and Gigolashvili, 1992; Sheeley, Wang, and Nash, 1992; Komm, Howard, and Harvey, 1993a, 1993b; Rybak, 1994; Ulrich and Bertello, 1996; Brajsa et al., 1997; Meunier, Nesme-Ribes, and Grosso, 1997; Brajsa et al., 1999, 2000; Howe et al., 2000a, b; Wohl and Schmidt, 2000; Antia and  Basu 2001; Howe et al., 2001; Brajsa et al., 2002; Stix, 2002; Altrock, 2003;   Heristchi and Mouradian, 2008). For a survey of the main results  of different
measurements of the solar differential rotation, please see the review papers of Howard (1984),
Schroter (1985), Snodgrass (1992),
Beck (1999), and Paterno (2010),  shown in which is a great achievement in the observations and analyses
of the solar surface differential rotation. Now,
it is generally believed that the solar differential rotation drives the solar dynamo for generating solar magnetic  activity (Babcock, 1961). However, the role of the differential rotation as a participant in
the cycle of magnetic activity variation is not yet clear. Hence,
the study of variations in the solar differential rotation is important
for understanding the Sun's internal dynamics and the
variations in the solar magnetic activity, as well as for finding
the cause of the variations in the differential rotation (Javaraiah, Bertello, and  Ulrich, 2005a; Chu et al., 2010).

Variation in the solar differential rotation is
an open question at present. For example, there are different and even contrary results for the long-term variation of solar rotation rate (Li et al., 2011a, b).
Lustig (1983), Howard (1984), Heristchi and Mouradian (2008), and  Li et al. (2011a, b) found that the general trend of the rotation rate of cycles should seemingly increase in the long run,
but the contrary result is obtained by Balthasar et al. (1986), Kitchatinov (1999), Zuccarello and Zappala (2003),
Brajsa et al. (2004), Javaraiah, Bertello, and  Ulrich (2005a),
 and Brajsa, Ruzdjak, and Wohl (2006).
In this study, we use the measurements of differential rotation given by Makarov,  Tlatov, and  Callebaut (1997),
Javaraiah, Bertello, and  Ulrich (2005a, b), and Javaraiah et al. (2009) to investigate
long-term evolution of solar differential rotation and the level of sunspot activity.

\section{Long-term Variations in the Solar Rotation and Activity}
\subsection{Long-term Variations of the Solar Differential Rotation}
The solar differential rotation can be determined
from the heliographic positions and the epochs of the observations
of large number of sunspots or sunspot groups using the
standard form:
$$
\omega (\phi)=A+B sin^{2}\phi
$$
where $\omega (\phi)$ is the solar
sidereal angular velocity at latitude $\phi$, and the coefficients $A$ and
$B$ represent the equatorial rotation rate and the latitudinal gradient
of the rotation, respectively (Howard, 1984; Javaraiah et al., 2003).
These two coefficients are inferred through a best fit to the data of observations.
Javaraiah, Bertello, and Ulrich (2005a) used the Greenwich data on sunspot groups during the period of 1879 January 1 $\,-\,$ 1976 December 31 and the spot group data from the Solar Optical Observing
Network (SOON) during the period of 1977 January 1 $\,-\,$ 2004
August 10 to determine the the coefficients $A$ and
$B$ in each of cycles 12 to 23, and the results were  given in the Table 1 of
Javaraiah, Bertello, and Ulrich (2005a). Based on the table, we plot here in Figure 1 the parameter $A$ of a solar cycle $vs$ the number of the cycle, respectively in the northern, the southern, and the two (both the northern and  southern) hemispheres. The unit of rotation rate originally in the table, $\mu rad\  s^{-1}$ is changed into
$deg\ day^{-1}$ by that $1\mu rad\ s^{-1}=4.95036\ deg\ day^{-1}$.
 A linear regression is taken between them respectively in the northern, the southern, and the two hemispheres, and resultantly, the correlation coefficients are -0.7713, -0.7811, and -0.8246, respectively.  These three correlation coefficients are all statistically significant at the $99\%$ confidential level. Thus, for the solar surface rotation rate at the Equator there exists a secular decrease of statistical significance. The secular slope is $-1.18\times10^{-2}\pm4.59\times10^{-3}$, $-1.00\times10^{-2}\pm3.85\times10^{-3}$, and  $-1.04\times10^{-2}\pm3.79\times10^{-3}$ ($deg\  day^{-1} cycle^{-1}$), respectively. If the period length of a solar cycle is approximately taken to be 11 years, then the secular slope is $-1.07\times10^{-3}\pm4.17\times10^{-4}$, $-9.01\times10^{-4}\pm3.50\times10^{-4}$, and  $-9.45\times10^{-4}\pm3.44\times10^{-4}$ ($deg\  day^{-1} year^{-1}$), respectively.

\begin{figure}    %%%%%%%%%%%%%%%%%% FIGURE 1
\centerline{\includegraphics[width=0.8\textwidth,clip=]{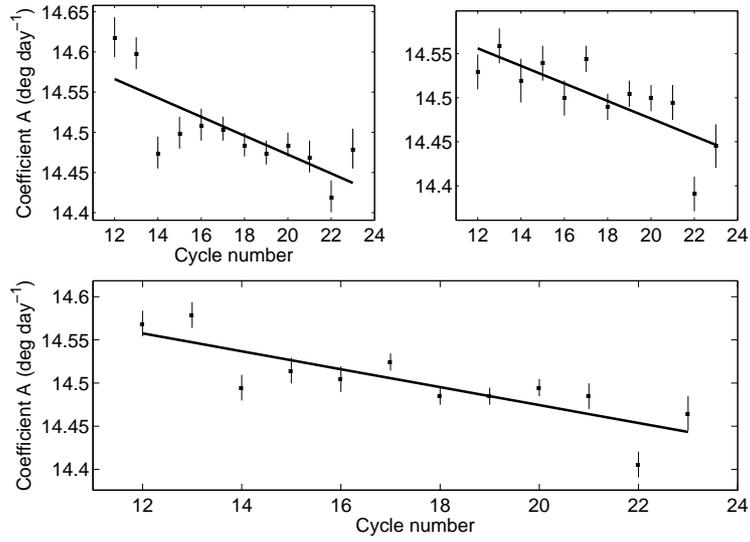}}
\caption{Cycle-to-cycle variations of coefficient $A$ (dots) respectively in the northern (the top left panel), the southern (the top right panel), and the two (the northern plus the southern) hemispheres (the bottom panel),  determined originally by Javaraiah, Bertello, and Ulrich (2005a) through using the Greenwich and SOON data of sunspots.
The solid lines are their corresponding linear regression trends.
The thin vertical lines are the corresponding error bars of coefficient A.
} \label{F-simple}
\end{figure}

The differential rotation of the large-scale magnetic fields was once investigated through studying the $H_{\alpha}$ synoptic charts during the years 1915 to 1990 by Makarov, Tlatov, and Callebaut (1997) and Kitchatinov et al. (1999). Based on the Figure 3 of Makarov,  Tlatov,  and Callebaut (1997), we plot here again in Figure 2 the time-dependent  rotation of the large-scale magnetic fields at the solar Equator. A linear regression is taken for the temporal rotation, and the correlation coefficient is -0.4945, which is statistically significant at the $99\%$ confidential level. Thus, for the solar surface angular velocity at the Equator there exists a secular decrease of statistical significance. The secular decrease is
$-1.48\times10^{-3}\pm3.84\times10^{-4}$($deg\  day^{-1} year^{-1}$).

\begin{figure}    %%%%%%%%%%%%%%%%%% FIGURE 2
\centerline{\includegraphics[width=0.8\textwidth,clip=]{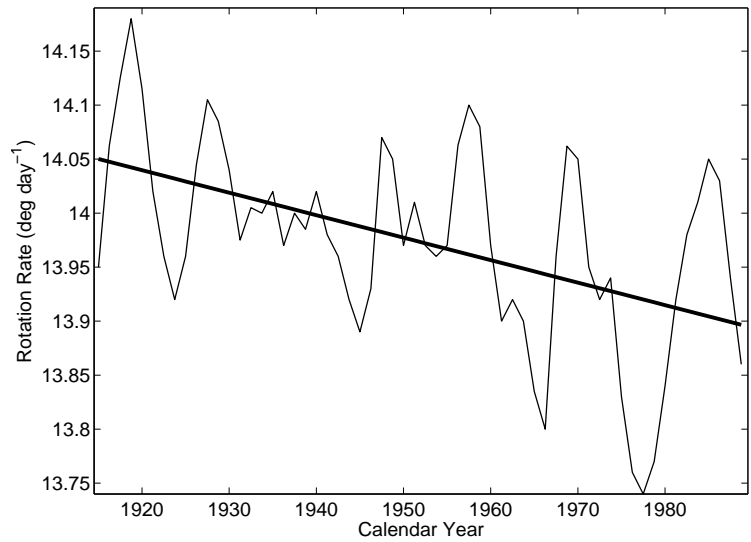}}
\caption{The temporal variation of the differential rotation of the large-scale magnetic fields at the solar Equator
 (the thin line), determined originally by Makarov,  Tlatov,  and Callebaut (1997). The thick line shows the corresponding linear regression trend.
} \label{F-simple}
\end{figure}

Javaraiah et al. (2009)  used the daily values of the equatorial rotation rate ($A$) derived from the
Mt. Wilson Doppler measurements during the period of 1985  December 3 to 2007 March 5, and they
 obtained the 61-day binned time series of the sidereal $A$ values. The 61-day interval
was chosen due to the existence of data gaps. Based on the Figure 3 of Javaraiah et al. (2009), we plot again here in Figure 3 the 61-day binned time series of the sidereal $A$ values. A linear regression is taken for the time series, and the correlation coefficient is -0.2533, which is statistically significant at the $99\%$ confidential level. Thus, for the solar surface rotation rate at the Equator there exists a secular decrease of statistical significance. The  decrease is $-1.38\times10^{-3}\pm4.98\times10^{-4}$($deg\  day^{-1} year^{-1}$).

\begin{figure}    %%%%%%%%%%%%%%%%%% FIGURE 3
\centerline{\includegraphics[width=0.8\textwidth,clip=]{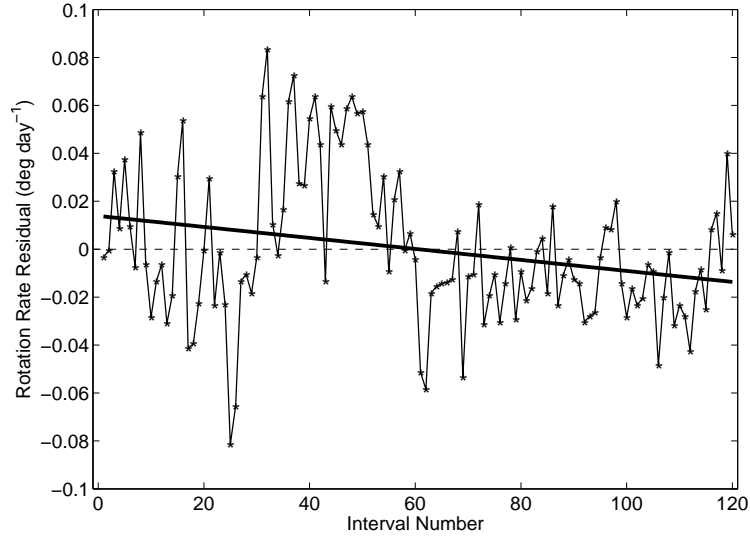}}
\caption{Variation of the equatorial rotation rate  in the
61-day intervals of the daily values derived from the
Mt. Wilson Doppler measurements during the period of 1985  December 3 to  2007  March 5
 (asterisks), determined originally by Javaraiah et al. (2009). The thin dashed horizontal line represents the mean value
 ($14.07 deg\ day^{-1}$), and the thick solid line shows  the corresponding linear regression trend.
} \label{F-simple}
\end{figure}

Although different methods and data are used, the aforesaid three studies all indicate that
for the solar surface rotation rate at the Equator there exists a secular decrease of statistical significance since cycle 12 onwards, and the secular decrease is about $1\,-\,1.5\times10^{-3}$($deg\  day^{-1} year^{-1}$).

Based on the Table 1 of Javaraiah, Bertello, and Ulrich  (2005a), we plot in Figure 4 the solar surface angular velocity of a  cycle $vs$
the number of the cycle, respectively in the northern, the southern, and the two hemispheres. The rotation angular velocity ($\overline{\omega (\phi)}$) of a cycle is taken as the average of $\omega (\phi)$ over latitudes of $0^{0}\,-\,40^{0}$. A linear regression is taken between them respectively in the northern, the southern, and the two hemispheres, and resultantly, the correlation coefficients are 0.3628, -0.2038, and 0.1083, correspondingly, which are all insignificant.
 If the rotation angular velocity ($\overline{\omega (\phi)}$) is taken as the average of $\omega (\phi)$ over latitudes of $0^{0}\,-\,50^{0}$, then the corresponding correlation coefficient is 0.3625, -0.2027, and 0.1095, respectively, with very slight changes, indicating that the taken latitude range should have a very slight influence.
Thus, for the solar surface rotation rate on an average of latitudes, there is not a secular trend of statistical significance.

\begin{figure}    %%%%%%%%%%%%%%%%%% FIGURE 4
\centerline{\includegraphics[width=0.8\textwidth,clip=]{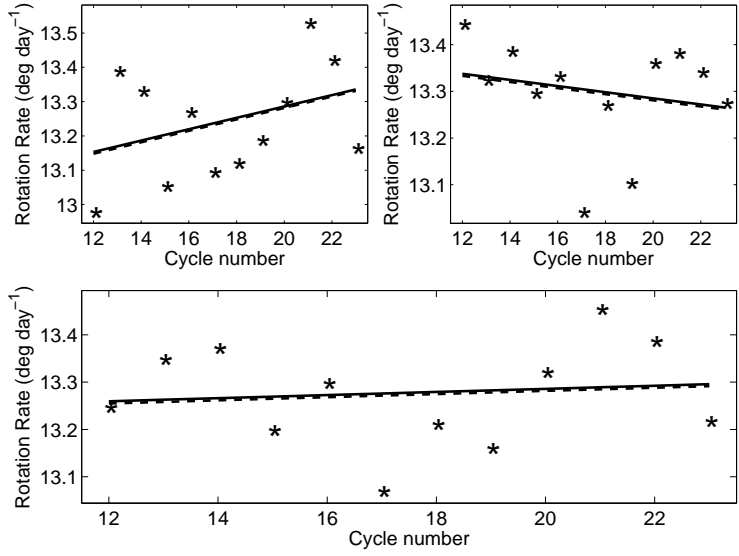}}
\caption{Cycle-to-cycle variations of the solar surface angular velocity (asterisks), respectively in the northern (the top left panel), the southern (the top right panel), and the two hemispheres (the bottom panel). The solid (dashed) lines are their corresponding linear regression trends, when the rotation angular velocity of a cycle is taken as the average of $\omega (\phi)$ over latitudes of $0^{0}\,-\,40^{0}$ ($0^{0}\,-\,50^{0}$).
} \label{F-simple}
\end{figure}

Based on the Table 1 of Javaraiah, Bertello, and Ulrich  (2005a), we plot in Figure 5  the parameter $B$ of a solar cycle $vs$
the number of the cycle, respectively in the northern, the southern, and the two hemispheres.
 A linear regression is taken between them respectively in the northern, the southern, and the two hemispheres, and resultantly, the correlation coefficients are 0.5301, 0.0936, and 0.3886, respectively.  The first correlation coefficient is statistically significant at the $92\%$ confidential level, but the later two are  insignificant.
 Thus, for the parameter $B$  there seems to exist a  secular decrease since cycle 12 onwards, but of weak statistical significance.

\begin{figure}    %%%%%%%%%%%%%%%%%% FIGURE 5
\centerline{\includegraphics[width=0.8\textwidth,clip=]{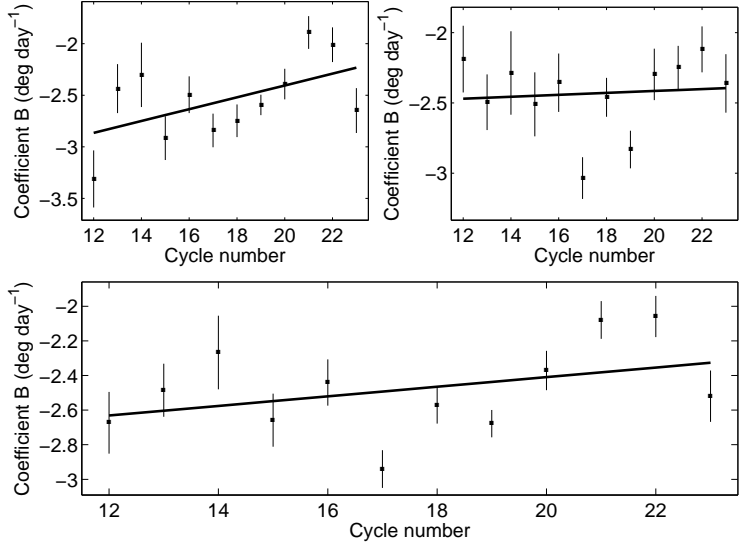}}
\caption{Cycle-to-cycle variations of  the coefficient $B$ (dots) respectively in the northern (the top left panel),  the southern (the top right panel), and the two  hemispheres (the bottom panel),  determined originally by Javaraiah, Bertello, and Ulrich (2005a) through using the Greenwich and SOON data of sunspots. The solid lines are their corresponding linear regression trends. The thin vertical lines are the corresponding error bars of coefficient B.
} \label{F-simple}
\end{figure}

\section{Relation between the Solar Rotation and the Level of Sunspot Activity in the Long Run}
The monthly sunspot areas
(in the sunspot area unit, namely, parts per million of the solar hemisphere)
respectively in the northern, the southern, and the two hemispheres can be obtained from
the NASA's website\footnote{http://solarscience.msfc.nasa.gov/greenwch.shtml}. They were complied by the Royal Greenwich Observatory during the period of the years 1874 to 1976 and by the Solar Optical Observing
Network of the US Air Force /US National Oceanic
and Atmospheric Administration  from 1977 onwards. We calculate the mean value ($\overline{Area}$) of the monthly areas over a solar cycle for each of cycles 12 to 23, respectively in the northern, the southern,  and the two hemispheres, which is shown in Figure 6.
A linear regression is taken between the mean value and the corresponding cycle number, respectively in the northern, the southern,  and the two hemispheres, and resultantly, the correlation coefficients are 0.6323, 0.6993, and 0.6851, respectively.   These three correlation coefficients are all statistically significant at the $96\%$ confidential level. Thus, for sunspot activity there exists a secular increase of statistical significance.

\begin{figure}    %%%%%%%%%%%%%%%%%% FIGURE 6
\centerline{\includegraphics[width=0.8\textwidth,clip=]{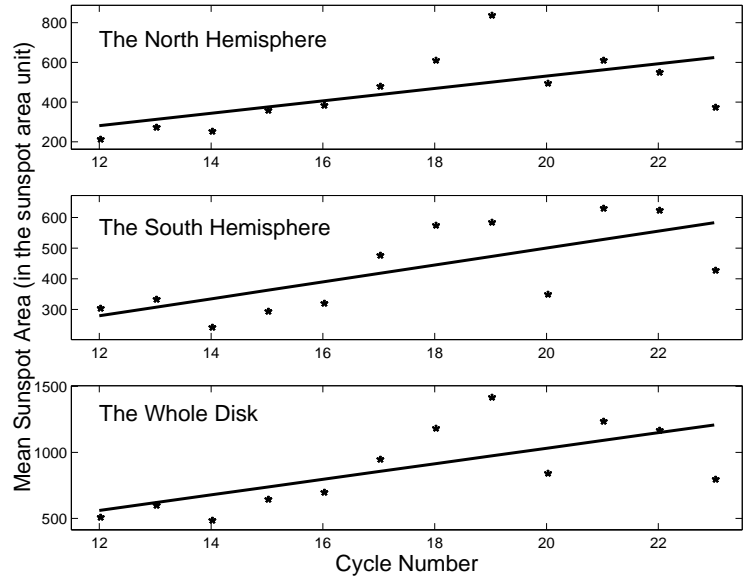}}
\caption{The mean value of the monthly areas (in the sunspot area unit) over a solar cycle for each of cycles 12 to 23 (asterisks), respectively in the northern (the top panel), the southern (the middle  panel), and the two hemispheres (the bottom panel). The solid lines are their corresponding linear regression trends.
} \label{F-simple}
\end{figure}

Based on the Table 1 of Javaraiah, Bertello, and Ulrich (2005a), we investigate the relation  of the mean value $\overline{Area}$ of the monthly sunspot areas over a solar cycle  $vs$ the parameter $A$ of the cycle for cycles 12 to 23, respectively in the northern, the southern, and the two hemispheres.
A linear regression is taken between them respectively in the northern, the southern, and the two hemispheres, and resultantly, the correlation coefficients are -0.5962, -0.5771, and -0.5881,  respectively.
All these three correlation coefficients are statistically significant at the $95\%$ confidential level. Thus, there exists a negative correlation between the level of sunspot activity, indicated
by the sunspot average areas and the solar equatorial rotation in the long run.
That is to say, the more rapidly the Sun rotates at the Equator, the lower sunspot activity level is.
%This is somewhat contrary to the solar dynamo theory.

%\begin{figure}    %%%%%%%%%%%%%%%%%% FIGURE 7
%\centerline{\includegraphics[width=0.8\textwidth,clip=]{figure7.eps}}
%\caption{Relation (dots) between the mean value $\overline{Area}$ of the monthly sunspot areas over a solar cycle  %$vs$ the parameter $A$ of the cycle determined originally by
%Javaraiah, Bertello, and Ulrich (2005a) through using the Greenwich and SOON data of sunspots,
%respectively in the northern (the bottom left panel), the southern (the bottom right panel), and the two hemispheres %(the up panel).  The solid  lines are their corresponding linear regression trends. The thin horizontal  lines are the %corresponding error bars of coefficient A.
%} \label{F-simple}
%\end{figure}

%We show in Figure 8 the relationship  of the yearly sunspot areas  $vs$ the parameter $A$ determined by Makarov,  %Tlatov,  and Callebaut (1997) and given in their Figure 3.
%A linear regression is taken for them, and the correlation coefficient is 0.1613, which is statistically %insignificant.

%\begin{figure}    %%%%%%%%%%%%%%%%%% FIGURE 8
%\centerline{\includegraphics[width=0.8\textwidth,clip=]{figure8.eps}}
%\caption{The relationship (asterisks) of the yearly sunspot areas  $vs$ the parameter $A$ determined by Makarov,  %Tlatov,  and Callebaut (1997) and given in their Figure 3.
% The solid  line is the corresponding linear regression trend.
%} \label{F-simple}
%\end{figure}

Based on the Table 1 of Javaraiah, Bertello, and Ulrich (2005a), we investigate the relation of the mean value $\overline{Area}$ of the monthly sunspot areas over a solar cycle  $vs$ the  parameter $B$ of the cycle for cycles 12 to 23, respectively in the northern, the southern, and the two hemispheres.
A linear regression is taken between them respectively in the northern, the southern, and the two hemispheres, and resultantly, the correlation coefficients are 0.3322,
-0.1378, and +0.1356,  respectively.
All these three correlation coefficients are statistically insignificant. Thus, there exists no relation between  the level of sunspot activity and the parameter $B$, viewed in the long run.

%\begin{figure}    %%%%%%%%%%%%%%%%%% FIGURE 8
%\centerline{\includegraphics[width=0.8\textwidth,clip=]{figure8.eps}}
%\caption{Relation (dots) between the mean value $\overline{Area}$ of the monthly sunspot areas over a solar cycle  %$vs$ the  parameter $B$ of the cycle determined using the Greenwich and SOON data of sunspots,
%respectively in the northern (the bottom left panel), the southern (the bottom right panel), and the two hemispheres %(the up panel).  The solid  lines are their corresponding linear regression trend. The thin horizontal  lines are the %corresponding error bars of coefficient B.
%} \label{F-simple}
%\end{figure}

Based on the Table 1 of Javaraiah, Bertello, and Ulrich (2005a), we plot in Figure 7 the mean value $\overline{Area}$ of the monthly sunspot areas over a solar cycle  $vs$
the solar surface angular velocity of the cycle, respectively in the northern, the southern, and the two hemispheres. The rotation angular velocity ($\overline{\omega (\phi)}$) is taken as the average of $\omega (\phi)$ over latitudes of $0^{0}\,-\,40^{0}$. A linear regression is taken between them respectively in the northern, the southern, and the two hemispheres, and resultantly, the correlation coefficients are 0.1897, -0.3710, and -0.0844, respectively. There exist both positive and negative coefficients, and further all these three values are statistically insignificant. Thus, there doesn't exist a relation between solar activity of a cycle and the surface rotation rate of the cycle in the long run. If the rotation angular velocity ($\overline{\omega (\phi)}$) is taken as the average of $\omega (\phi)$ over latitudes of $0^{0}\,-\,50^{0}$, then the corresponding correlation coefficients are 0.1904, -0.3701, and -0.0835, respectively, with very slight changes, indicating that the taken latitude range should have a very slight influence.

\begin{figure}    %%%%%%%%%%%%%%%%%% FIGURE 7
\centerline{\includegraphics[width=0.8\textwidth,clip=]{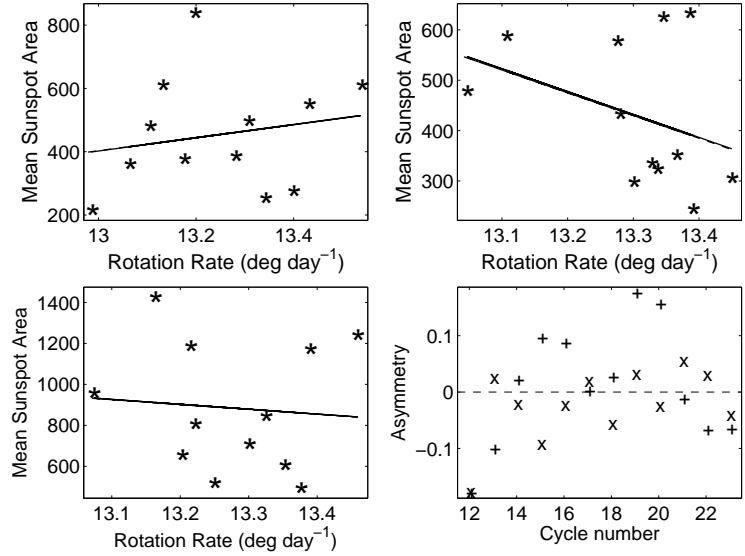}}
\caption{The mean value $\overline{Area}$ of the monthly sunspot areas over a solar cycle  $vs$
the solar surface angular velocity of the cycle, respectively in the northern (the top left panel), the southern (the top right panel), and the two hemispheres (the bottom left panel). The solid lines show their corresponding linear regression. The bottom right panel:  the north-south asymmetry values of sunspot area (plus symbols) and the solar surface angular velocity (cross symbols) in cycles 12 to 23.
} \label{F-simple}
\end{figure}

Solar-activity indexes vary over the solar disk, and various activity indexes cannot be considered
to be symmetrical between the solar northern ($N$) and southern ($S$) hemispheres, it is the so-called
 $N\,-\,S$ asymmetry of solar activity (Chang 2008, 2009). To characterize the $N\,-\,S$ asymmetry of solar activity, it is traditionally normalized to the solar
activity present in both the solar hemispheres and defined as follows:
$Asymmetry = (NO_{N}- NO_{S})/(NO_{N} + NO_{S})$, where $NO_{N}$ and $NO_{S}$ stand for the values of the solar
activity indicators considered corresponding to the northern and southern hemispheres,
respectively. The $N\,-\,S$ asymmetry of sunspot area in each of cycles 12 to 23 was given by Li et al. (2002), and
Li et al. (2009), which is also shown here in Figure 7. Based on the table of Javaraiah, Bertello, and Ulrich (2005a), the $N\,-\,S$ asymmetry of
the average solar rotation angular velocity in each of cycles 12 to 23 is calculated and shown in the figure. Here,
The rotation angular velocity ($\overline{\omega (\phi)}$) is taken as the average of $\omega (\phi)$ over latitudes of $0^{0}\,-\,40^{0}$.
If solar activity of a cycle is statistically related to the surface rotation rate of the cycle, their asymmetry values should statistically have the same  sign when the two have the same secular trend, or the opposite sign when the two have the
opposite trend. As the figure shows, sunspot areas obviously have the same  sign of asymmetry values as surface rotation rates have in three cycles, cycles 12, 19,  and 23, while clearly having the opposite sign in eight cycles, cycles 13, 14, 14, 16, 18, 20, 21, and 22. Thus, there seems no correlation between the asymmetries in sunspot average area and the average rotation rate in the period analyzed.
If the rotation angular velocity ($\overline{\omega (\phi)}$) is taken as the average of $\omega (\phi)$ over latitudes of $0^{0}\,-\,50^{0}$, then the same sign of the $N\,-\,S$ asymmetry is obtained as when the average of $\omega (\phi)$ is over latitudes of $0^{0}\,-\,40^{0}$.

\section{Conclusion and Discussion}
Javaraiah, Bertello, and Ulrich (2005a) used the Greenwich data on sunspot groups during the period of 1879 January 1 $\,-\,$ 1976 December 31 and the spot group data from the Solar Optical Observing
Network (SOON) during the period of 1977 January 1 $\,-\,$ 2004
August 10 to determine the coefficients $A$ and
$B$ in the standard form of the differential rotation. They studied long-term variations in solar differential rotation and sunspot activity. Makarov,  Tlatov, and  Callebaut (1997) investigated the differential rotation of the large-scale magnetic fields through  the $H_{\alpha}$ synoptic charts during the years 1915 to 1990.
Javaraiah et al. (2009) used the daily values of the equatorial rotation rate ($A$) derived from the
Mt. Wilson Doppler measurements during the period of 1985  December 3 to 2007  March 5. They
 obtained the 61-day binned time series of the sidereal $A$ values.
In this study, we use the above three kinds of data to revisit the relationship between the
long-term variations of solar differential rotation and the level of sunspot activity.

Although different methods and data are used, the aforesaid three studies all indicate that
for the solar surface rotation rate at the solar Equator there exists a secular decrease of statistical significance since cycle 12 onwards, and the secular decrease is about $1\,--\,1.5\times10^{-3}$($deg\  day^{-1} year^{-1}$).

The parameter $B$ represents the differentiation degree of the surface differential rotation: the larger the absolute $B$ is, the more pronounced the differential rotation is.
For the parameter $B$  there seems to also exist a  secular decrease since cycle 12 onwards, but of weak statistical significance. For the solar surface rotation rate on an average of latitudes, there is not a secular trend of statistical significance.
That is to say, the rotation angular velocity is changed from having a significant secular trend  to having no secular trend, when latitudes are considered from the  Equator to higher latitudes.
%Secular trend of the rotation angular velocity  on an average of latitudes or at a certain latitude should change with %latitudes.
It is known that sunspots with different property, such as size, life time and so on, appear in different phase of a solar cycle and in different latitudes. Thus, the observed changes of rotation by various tracers couldn't represent a global variation of the solar rotation, that is to say, they are caused by some specific property of the tracer used.
This is probably the reason why different secular trends are found by different data and different methods.
And as pointed out by Brajsa, Ruzdjak, and Wohl (2006), {\it the spatial-temporal interplay can often be hidden in the results by various tracers used, and a straightforward comparison of different results is quite difficult}.

We calculate the mean value  of the monthly areas over a solar cycle for each of cycles 12 to 23, and find that  it has a secular increase of statistical significance. For the cycle-to-cycle variations of sunspot activity and the solar equatorial rotation, there is a negative correlation between the two in the long run.
The more rapidly the Sun rotates at the Equator, the lower sunspot activity level is.
%This is somewhat contrary to the solar dynamo theory.

Our results show no relation between  the level of sunspot activity and the parameter $B$, viewed in the long run, and
there doesn't exist a relation between solar activity level of a cycle and the surface rotation rate of the cycle in the long run.
%That is to say, strong magnetic fields do not lead to a small absolute value of the parameter B.
%Thus, it is impossible to expect a more pronounced differential rotation yielding a higher rotation velocity at low %latitudes on an average, when the solar magnetic fields are weaker.

%%%%%%%%%%%%%%%%%%%%%%%%%%%%%%%%%%%%%%%%%%%%%%%%%%%%%%%%%%%%%%%%%%%%%%%%%%%
\begin{acks}
We thank the anonymous referees for their careful reading of the
manuscript and constructive comments which improved the original
version of the manuscript.
This work is supported by the
Natural Science Funds of China (10873032, 10921303, 11147125, and 11073010), the 973
programs 2011CB811406 and 2012CB957801, and the Chinese Academy of Sciences.
\end{acks}

%%% BIBLIOGRAPHY %%%%%%%%%%%%%%%%%%%%%%%%%%%%%%%%%%%%%%%%%%%%%%%%%%%%%%%%%%%
\mbox{}~\\
\bibliographystyle{spr-mp-sola-cnd} %% Alternative style: no title,
                                      % no concluding page.
\IfFileExists{\jobname.bbl}{} {\typeout{}
\typeout{****************************************************}
\typeout{****************************************************}
\typeout{** Please run "bibtex \jobname" to obtain} \typeout{**
the bibliography and then re-run LaTeX} \typeout{** twice to fix
the references !}
\typeout{****************************************************}
\typeout{****************************************************}
\typeout{}}

\end{article}
\end{document}